\documentclass[namedreferences]{SolarPhysics}
\usepackage[optionalrh]{spr-sola-addons} 
\usepackage{graphicx}        
\usepackage{color}           
\usepackage{url}             

\newcommand{\Rsun}{{R_{\odot}}}
\newcommand{\mydeg}{{^{\circ}}}
\newcommand{\kms}{{km s^{-1}}}
\newcommand{\agu}{  {\it AGU}}

\newcommand{\antok}{  {\it Ann. Tokyo Astr. Obs.}}
\newcommand{\araa}{   {\it Ann. Rev. Astr. Astrophys. Suppl.}}

\newcommand{\aap}{    {\it Astron. Astrophys.}}

\newcommand{\apj}{    {\it Astrophys. J.}}

\newcommand{\sci}{    {\it Science}}
\newcommand{\solphys}{{\it Solar Phys.}}

\newcommand{\ssr}{    {\it Space Sci. Rev.}}

\begin{document}

\begin{article}

\begin{opening}

\title{Hybrid Reconstruction to Derive 3D Height-Time Evolution for Coronal Mass Ejections}

\author{A.~\surname{Antunes}$^{1},^{2}$
        A.~\surname{Thernisien}$^{1},^{3}$
        A.~\surname{Yahil}$^{4}$
       }
\runningauthor{Antunes et al.}
\runningtitle{H-T Evolution via Hybrid Reconstruction}

   \institute{$^{1}$ Naval Research Laboratory
                     email: \url{alexander.antunes@nrl.navy.mil}\\ 
              $^{2}$ National Research Council\\
              $^{3}$ USRA\\
              $^{4}$ ImageRecon LLC\\
             }

\begin{abstract}

We present a hybrid combination of forward and inverse reconstruction methods using multiple observations of a coronal mass ejection (CME) to derive the 3D 'true' Height-Time plots for individual CME components.
We apply this hybrid method to the components of the 31 Dec 2007 CME.
This CME, observed clearly in both the STEREO A and STEREO B COR2 white light coronagraphs, evolves asymmetrically across the 15 solar radius field of view within a span of three hours.  The method has two reconstruction steps. We fit a boundary envelope for the potential 3D CME shape using a flux rope-type model oriented to best match the observations.  Using this forward model as a constraining envelope, we then run an inverse reconstruction solving for the simplest underlying 3D electron density distribution that can, when rendered, reproduce the observed coronagraph data frames.  We produce plots for each segment to establish the 3D or ``true'' Height-Time plots for each center of mass as well as for the bulk CME motion, and use these plots along with our derived density profiles to estimate the CME asymmetric expansion rate. 
\end{abstract}
\keywords{Coronal Mass Ejections, Initiation and Propagation}
\end{opening}

\section{Introduction}
     \label{S-Introduction} 

3-Dimensional (3D) Reconstruction is one method of stereoscopic analysis and involves
the process of using multiple views of a single object from different
view angles in an effort to determine the intrinsic 3D
shape of the object.  With one viewpoint, only 2-dimensional feature
qualities are observed and any underlying 3D structure must be
inferred from theory.  With more than one viewpoint, comparison of the
projection effects in the images can be used to mathematically invert
the images to determine an optimal 3D distribution of matter that
would best reproduce the observed images, such as with rotational
tomography \cite{Frazin05} to provide multiple viewpoints over time.

Coronal Mass Ejections (CMEs) are rapidly evolving, fast moving heliospheric
plasma ejecta originating in the lower solar atmosphere.  Early LASCO
\cite{Brueckner95}
and EIT \cite{Dela95} results included the first observations of the start of a CME
\cite{Wood99} and analysis included comparisons against the
'flux rope' model.  In one sample CME
event, a prominence accelerated to 100$km s^{-1}$ and then, within the view
range of 1.1 to 30 solar radii in the corona, accelerated to 200-400
km/s \cite{Dere97}, ultimately covering 70 degrees of visual
latitude.  CMEs which are directed towards Earth can cause disruptive
effects when they hit and interact with the Earth's magnetic field.

To model dynamic features
such as Coronal Mass Ejections (CMEs), the multiple view angles must also be simultaneous or
near-simultaneous.  With the two view angles from Solar Terrestrial Relations Observatory (STEREO, \opencite{Kaiser08}) A and B satellites, optionally combined with additional viewpoints from other satellites, 3D
reconstruction of rapidly evolving features such as CME is possible
\cite{Cook02} but the problem is underconstrained when there are too
few viewpoints.

Forward modeling (FM) presumes a theoretical shape for a CME and seeks to
position that shape so as to reproduce the observed data.  For data
with strong asymmetries, additional model assumptions must be added
to accurately reproduce the data.  Inverse modeling makes no shape
assumptions when 'carving' out an underlying density distribution that, when
rendered, reproduces the observed data.  For underconstrained problems,
there are multiple possible density distributions that can reproduce
the same data, so some criteria for weighting potential solutions must
be employed.

Since CME structure is neither perfectly formed like a model, nor
completely unknown as inversion assumes, we combine the two methods
to make minimal assumptions on shape and extent as a constraint
for the model-free inverse solution.

\begin{figure}
\centerline{
\includegraphics[width=0.515\textwidth,clip=]{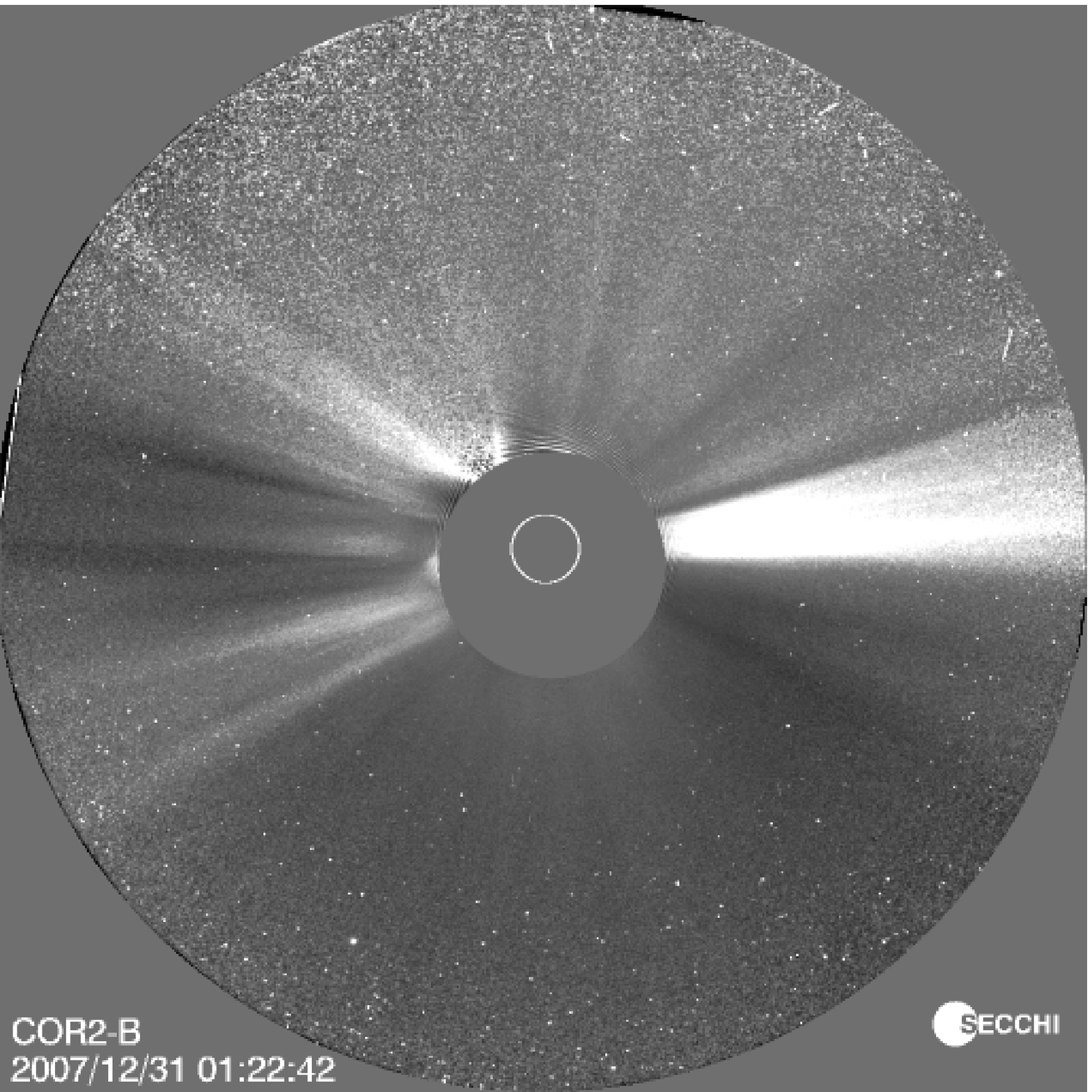}
\includegraphics[width=0.515\textwidth,clip=]{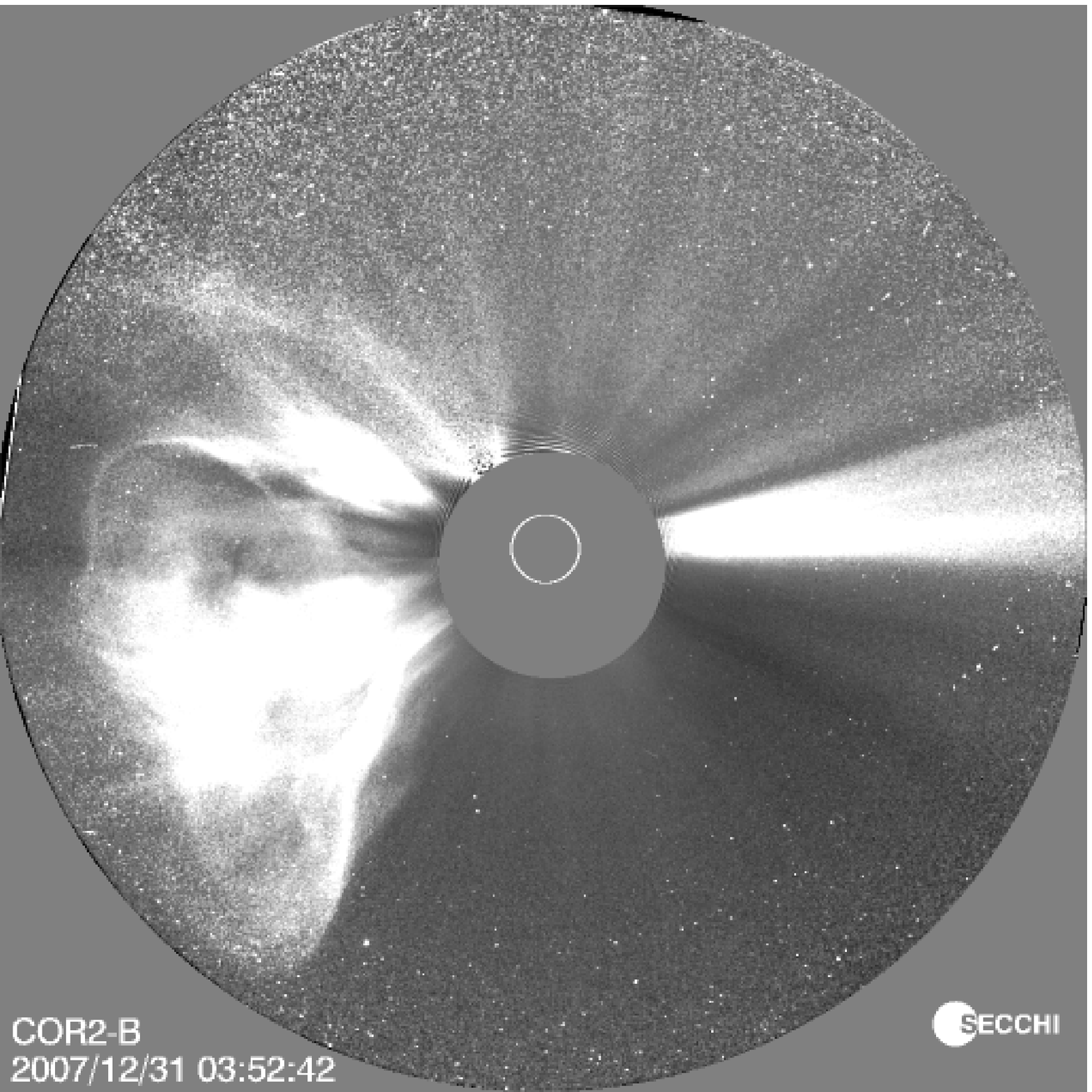}
}
\centerline{
\includegraphics[width=0.515\textwidth,clip=]{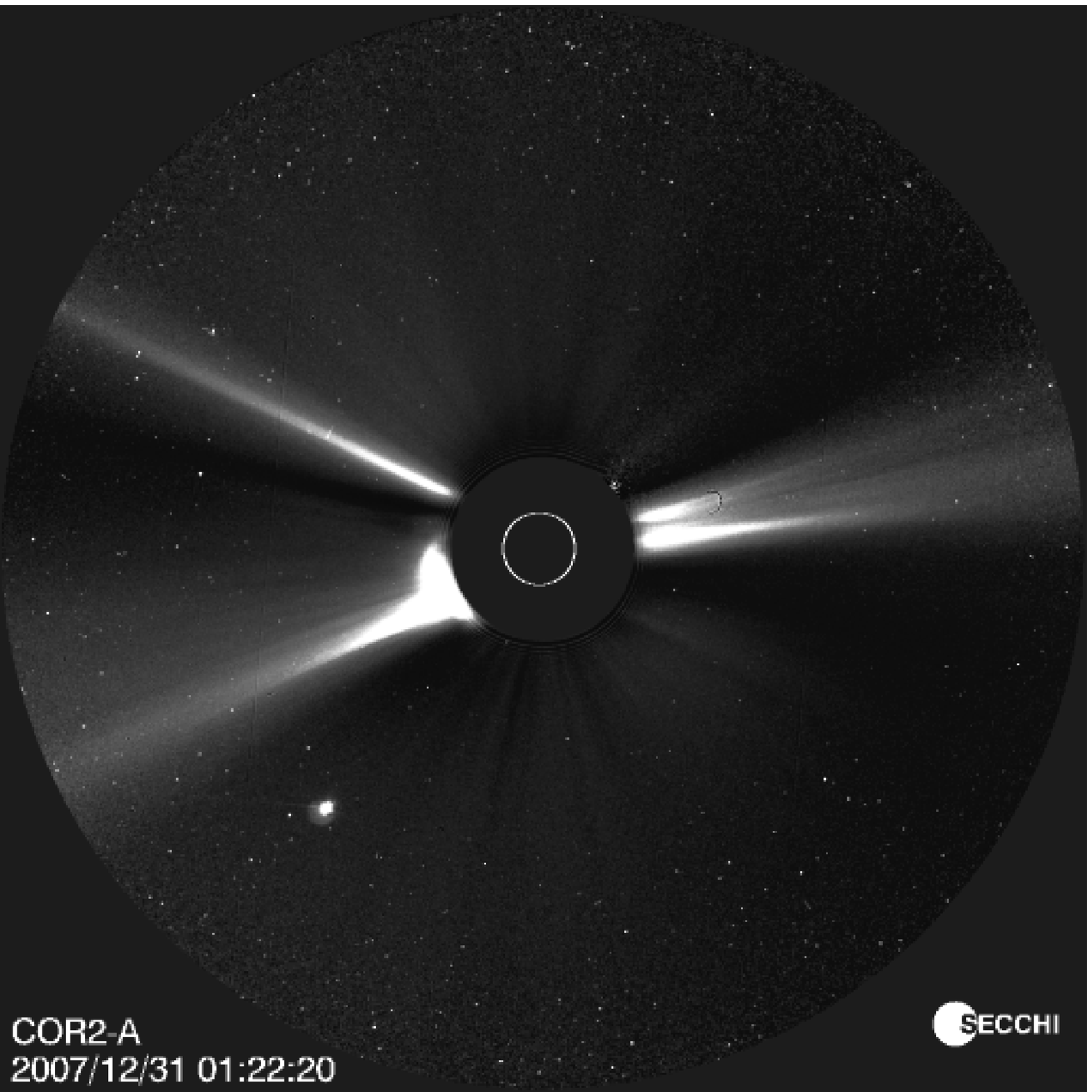}
\includegraphics[width=0.515\textwidth,clip=]{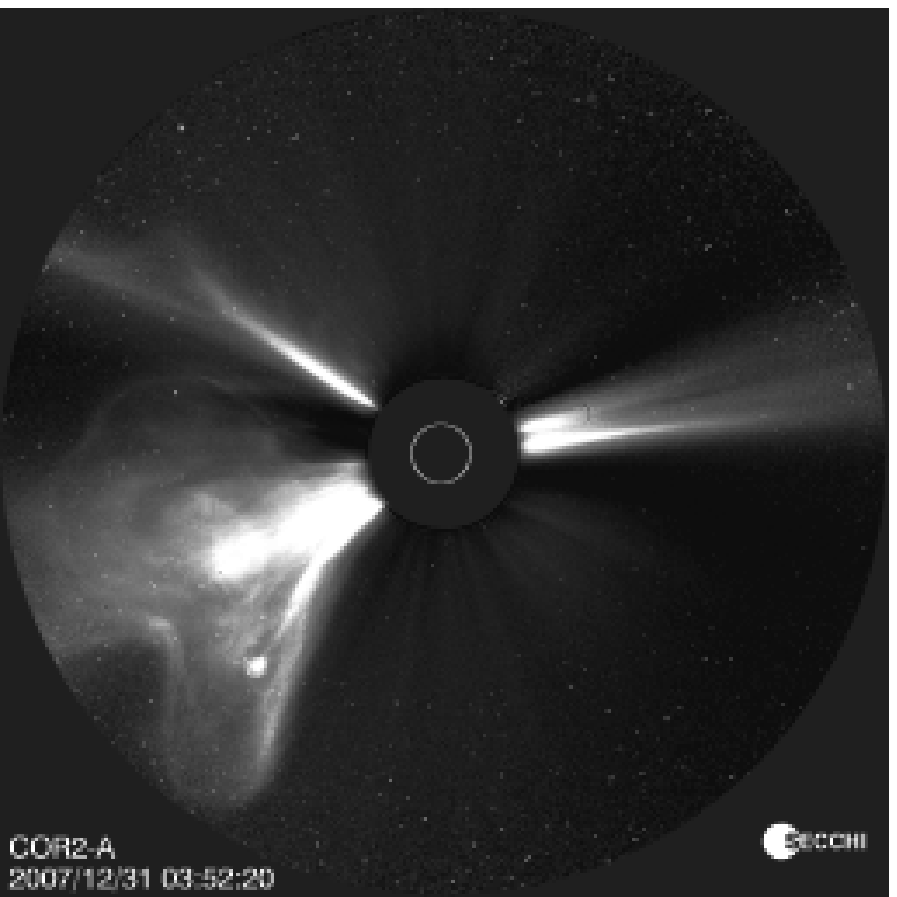}
}
\caption{The 31 Dec 2007 CME seen from COR2-B (upper panels) and COR2-A (lower panels) at a very early state at 01:22UT (left panels) and as it approaches the edge of the field of view at 03:52UT (right panels).  These images are processed using a weighted monthly median background subtraction to remove quasi-stable corona features while maintaining streamers, CMEs and transients. Our reconstruction, in contrast, uses difference images to remove the streamers and quasi-steady transients to best view only the rapidly-changing CME.  These images are oriented such that solar north coincides with the y-axis of the image.}
\label{FigureCME}
\end{figure}

In our hybrid reconstruction, we define the detailed 3D density distribution for the CME and derive the 3D Height-Time from the Sun for each data frame.  This is a true 3D position for the chosen center of mass with no projection effects.  Prediction of the arrival of a CME to 1AU is strongly correlated with true space speed and the subsequent deduction of initial acceleration/deceleration \cite{Michalek04}.  From the reconstruction time series of true 3D positions, we can derive a true space speed.  

\section{Observations of the 31 December 2007 CME}
     \label{S-CME}

\subsection{STEREO/SECCHI Instrumentation}

Our primary source of data is from 
the Sun-Earth Connection Coronal and Heliospheric Investigation (SECCHI, \opencite{Howard08})
instruments on board STEREO, as
these return synchronized images from two viewpoints using multiple
detectors.  STEREO is a pair of identical spacecraft, A and B,
launched into a leading and trailing Earth orbit.  The twin
spacecraft, equipped with identical instruments, observe the Sun
simultaneously from two different angles, allow for the first time
3D viewing of solar features such as CMEs.
The initial fields of view are partially
overlapping and, for the 31 Dec 2007 event, the satellites are at an angular spacing of 44 degrees.

The COR2 white light coronagraphs onboard each of the twin STEREO spacecraft each
cover a field of view from 2 to 15 solar radii.
All SECCHI COR2 images used are total brightness images
acquired with an cadence of every 30 minutes.

\subsection{31 December 2007 CME}

We are primarily concerned with the density and kinematics of the 31 Dec 2007 CME as derived from the COR2 observations, however, we briefly look at this CME in context.  There is a pre-existing streamer, visible when median background subtraction is used (Figure \ref{FigureCME}) but not visible in difference images, as it changes only slowly relative to the COR2 cadence.
A CME erupts from this active region, either as a streamer blow-out type event, a prominence eruption, or an impulsive lower event.  This temporarily changes the magnetic field topology and either completely dissipates the streamer or minimally deflects it. After the event, the basic magnetic field structure reforms and subsequently, the streamer reforms.

\begin{figure}
\centerline{\includegraphics[width=0.9\textwidth,clip=]{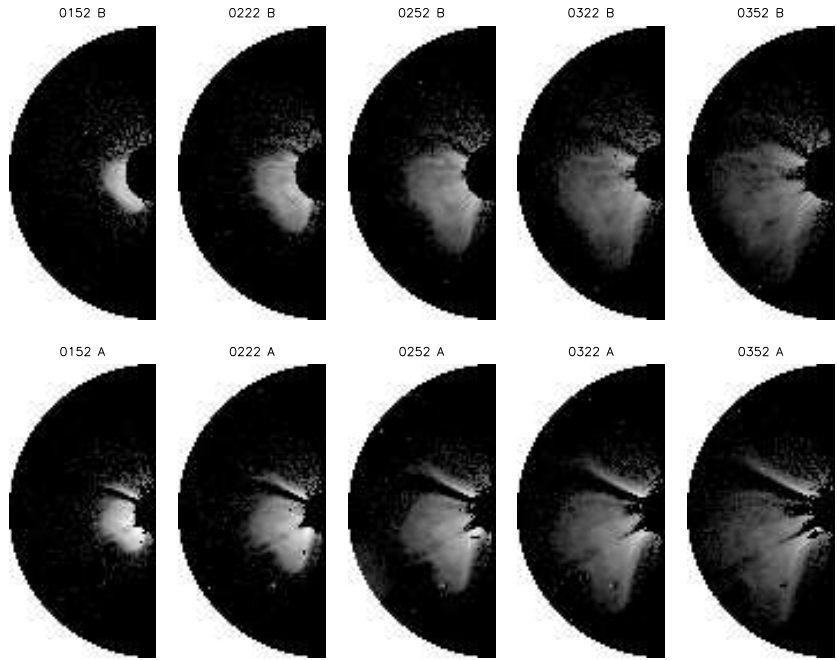}}
\caption{The 31 Dec 2007 CME as seen by COR2-B (upper panels) and COR2-A (lower panels) from 1:52am to 3:52am (time increasing left to right).  We have masked the non-CME portions, including the non-CME half of each detector and outer corners, in order to isolate and emphasize the CME itself.  Plots are base difference images subtracting a pre-CME frame at 0:22 from the data to extract the fast-moving CME from background and quasi-stable structures.  We unfortunately note the presence of a non-CME instrument artifact in data frame COR2-A 02:52.}
\label{FigureEvo}
\end{figure}

CMEs are faint relative to the underlying F- and K-corona emission, so
we use difference images to extract the CME from the background (Figure \ref{FigureEvo}).  We subtract the 0:22UT 'quiet' data frame from each image in order to better view the CME.  This subtracts the F-Corona as well as any quasi-static non-CME structures.
We confine the kinematic analysis to the period during which the CME is visible in the COR2 field of view.  By the time the CME has entered the COR2 field of view, the CME front has become distorted and shows a distinct upper and lower lobe (again, Figure \ref{FigureCME}).

The CME event was first visible within the COR2 coronagraphs
at around 1:52UT, and the leading edge begins to depart at or around 4:52UT.
We use base difference images as explained above, shown in
Figure \ref{FigureEvo}. We also create a running difference image consisting of a data frame minus the immediately prior (by 30 minutes) data frame, to extract just the most rapidly changing portion of the CME.

\subsection{CME Kinematics}
	\label{S-Kinematics}

There is not yet one single 'best' method in the fledgling field of solar reconstruction \cite{Newmark04}.  In addition to our choice of reconstruction method, we note that other methods are used to estimate the location of a CME.
Image geometric constraints can localize a CME rapidly \cite{Mierla08} and, if the CME is fully in the field of view, deduce the total mass and center of mass \cite{Colaninno09}.
Polarization measurements allow 3D placement using a
single satellite (\opencite{Moran04}, \opencite{Dere05}).
Methods such as tie-pointing and visual analysis \cite{Stenborg08} manually draw where the presumed shock front or leading edge of the CME is, then use that to define the Height-Time profile for the CME.

While the center of mass of a CME in a single image can be computed \cite{Vourlidas00}, this center of mass cannot be directly compared with that from a different satellite image unless it is known that the full CME mass is visible in both.
Further, you cannot define identical components, such as the upper and lower lobes seen in this 31-Dec-2007 CME, in two views (such as COR2-A and COR2-B) unless the CME is both equally visible in both detectors and perfectly aligned.  If the CME has any tilt or asymmetry, material from one region in COR2-A will be co-mingled with material viewed in COR2-B.  The solution is to reconstruct the entire underlying density, then split that 3D distribution into the appropriately-oriented regions to extract component masses and kinematics.

With STEREO, we have two simultaneous viewpoints of CMEs and solar features at wide angles of separation that do not necessarily show the same morphological shape in all viewpoints, in which the CME may not be fully within the field of view of all viewpoints at all times.
We use a hybrid inverse reconstruction method to fit the underlying CME density distribution without requiring distinct identifiable features nor mass conservation.
This hybrid method is a type of full-CME modeling that solves for the most likely underlying density distribution that can give rise to the data observed.  From the resulting 3D solution, we can then produce total mass and regional section masses as well as reconstructing successive frames to obtain 3D kinematic information.
We solve for the bulk CME, generating results that differ from that of leading edge and shock front methods.  We also provide an equivalent for the leading edge, where we trace the center of mass of the most rapidly changing material through a running difference method.

\section{Hybrid Reconstruction}
	\label{S-HybridRecon}

\subsection{Reconstruction Methods}
	\label{S-Recon}

A 3D reconstruction problem consists of selecting data, extracting the geometry of the view, rendering the data, then applying a reconstruction method.
Inverse modeling is a non-parametric method that assumes no a priori shape for the CME and uses grid minimization to solve for an underlying 3D electron density distribution by comparing renderings with the actual data views taken. Inverse modeling is constrained by the limits on the overlapping viewpoints and cannot provide model insight into regions that do not contain at least two partially overlapping fields of view. In our work we use the two STEREO viewpoints, in contrast to wide-field image work \cite{Jackson02} and rotational tomography \cite{Frazin05}. In particular, the existing data are all along the ecliptic and lack an 'overhead' view from the pole.

In contrast to inversion methods, which are non-parametric and produce non-unique solutions, forward modeling assumes a specific parametric shape and fits the model until it produces renderings that match the actual data. Forward modeling therefore provides a likelihood that a given model shape describes the actual CME event, but does not provide information on other models distributions that could produce the same data.

In essence, inverse modeling is underconstrained and forward model is overconstrained.  Given enough viewpoints, an inverse modeling solution will produce the best match for the actual underlying distribution. In regions where the solution space is under-resolved and has insufficient viewpoints, inverse modeling often produces a non-physical solution. Forward modeling produces a physical solution based on model assumptions, but predisposes the solution to only fitting that model.

\subsection{Combining Inversion with Forward Modeling}
	\label{S-AppMethod}

Our
hybrid method combines model-free inverse reconstruction supplemented with weighted model renderings for regions where the inverse reconstruction problem is unresolved.  This balances model-centric approaches with the unbiased solution from inverse modeling.  By creating synthetic model viewpoints at a lower weight than the actual data, the inverse reconstruction can better converge on a physically meaningful result without being overly predisposed towards a single model assumption.

The inversion computes a presumed underlying density distribution in an attempt to reproduce the data.  The data need to be fit only to the tolerance set by the noise, since COR2-A and COR2-B views will have statistical deviations as well as instrument effects.
For mass that is visible in one view but not another (such as being behind the occulter in one view), that portion of the inversion solution is less constrained by lacking the second view, but will still place mass that does not violate either data view.  As the CME evolves into and out of each field of view, there will be uncertainties in the mass due to fewer constraints, but the solutions themselves do not assume equal mass visible.  The final inversion result will be self-consistent with all available data, though it does not require that the solution be unique.

We use this hybrid of inverse and forward
modeling to
investigate the asymmetries of the 31 Dec 2007 CME as observed by the SECCHI
COR2 white light coronagraphs onboard the twin STEREO satellites.
We observe a streamer and the CME superimposed in the image (Figure \ref{FigureCME}) and see evidence for interaction of the two.  We bisect the CME into an 'upper' and 'lower' region using the line defined by the movement of the bulk center of mass radially out from the Sun.  This vector also corresponds roughly to the observed longitude and latitude of the streamer seen in the COR2 images.

By combining forward and inverse modeling into a single hybrid method, we reconstruct a probable underlying 3D density distribution for each pair of frames at a given time.  From this, we can sum the mass to obtain a total mass and center of mass position, or divide the mass to compare different 3D regions of the distribution.  We compare the total mass (and total CM) to that of the distinct upper and lower regions.  By comparing the distribution over time, we obtain the kinematics for the CME as a whole as well as for the bisecting regions.

\subsection{K-Corona}
	\label{S-Kcorona}

The visible (white light) emission of the solar K-corona is due to Thomson scattering of the incident photospheric emission by hot coronal electrons. This scattered radiation is  tangentially and radially polarized.  SECCHI coronagraphs mask the solar disk, whose brightness is more than $10^{5}$ that of the corona.  The coronagraph measures the brightness integrated over the line of sight through the optically thin corona.  Applying G(x,y,z), a geometric function for the line of sight (depending on limb darkening) yields, for a given distance 'z' to the line of sight, the following total brightness \cite{Billings66}:

\begin{equation}
tB = C \int  N_{e}(x,y,z) G(x,y,z) dz
\end{equation}

Rendering of a presumed underlying density distribution is generally solved using raytracing, where we sum the Thomson scattering contribution of each 3D 'voxel' element stepwise along the line of sight to obtain the total brightness at the given detector pixel.

We use difference images in order to subtract the F-Corona and quasi-stable structures from the data.  The data that remains is presumed to be entirely K-corona white light emission due to the rapidly CME changing material.

\subsection{Application of the Method to the Data}
	\label{S-AppData}

To reconstruct, we first fit the CME loosely using forward modeling \cite{Thernisien09}.  A flux rope shaped density distribution is manually positioned until its shape overlays with the observed shape in both the COR2-A and COR2-B data frames.  The model assumes a symmetric CME and serves as only a rough fit to the actual CME.  This model produces a likely enclosing volume for the CME, but does not provide us with information on mass or density.

\begin{figure}
\centerline{\includegraphics[width=0.6\textwidth,clip=]{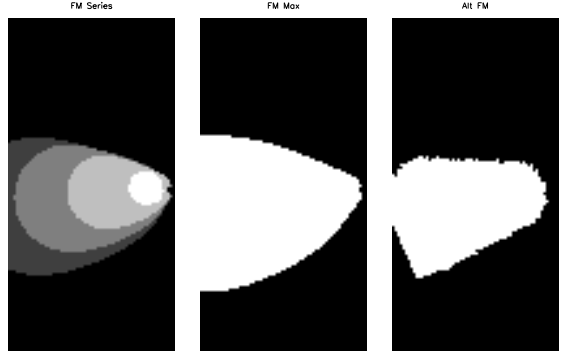}}
\caption{Top-down (polar) view of the choice of constraining envelopes used to restrict the reconstruction.  Three different Forward Model assumptions were tested.  'FM series' (left panel) uses the Thernisien FM for each time step.  'FM max' (middle panel) applies only the largest FM to each step.  'Alt FM' (right panel) uses a different maximum FM envelope.  Our reconstructions use the middle panel 'FM max' in producing these numerical results.}
\label{FigureFMs}
\end{figure}

Second, we use this forward model to generate a simulated polar view of the CME (Figure \ref{FigureFMs}).  From this, we create a polar mask to serve as a constraining envelope.  Our subsequent modeling will thus require that the CME be fully contained within this 3D cylindrical envelope.

Finally, we use the COR2-A and COR2-B data with the addition of the constraining polar mask to run an inverse reconstruction.  The inversion will place density within the enveloped region in a manner which, when rendered, will recreate the original observed data.  This allows the inversion to capture any asymmetries within the actual density distribution while preventing the inversion from placing material outside what theory (via the forward model) suggests is the constrained volume of space for the CME.

Our forward modeling technique converges on parameters of a given geometric model-- a 'croissant' flux rope shell model \cite{Thernisien09}-- to find the best fit that model gives to the original data.  These fits are done by 'eye' to place a flux rope that best matches the morphology seen by the multiple data views.
For inverse modeling, we converge on a parameter-free minimization of the underlying matter distribution.  Using a cube (consisting of 'voxels', the 3D equivalent to 2D pixels), we first render a possible solution volume, compare that render to the original data, then adjust our guess at the solution iteratively until the renders match the data.

Because there are multiple solutions to the underlying distribution that can potentially reproduce the original data, we try to create the simplest underlying density distribution which can reproduce the original data.
We currently use the PIXON method, discussed in detail in \opencite{Puetter05}.  PIXON is a minimum complexity, non-parametric, locally adaptive, iterative image reconstruction method.  It seeks the smoothest solution that fits the data, using adaptive elements dubbed 'Pixons' to restrict the number of elements needed for the calculation.  The Pixon elements smooth the data to reduce the number of image components.  Given an initial image, PIXON finds a non-negative least-squares image, maps that image with Pixon elements, then updates the solution image with the Pixon map.  A solution with fewer Pixon elements is deemed simpler and therefore more likely.

Since the data are all gathered from the ecliptic plane, the reconstructions are not well constrained along the polar viewpoint.  Using a forward model, we can create an artificial estimate at a polar view, and use that to restrict the spatial extent of the possible solution.
So we further constrain our inverse reconstruction by using a forward model of the CME to create an outer boundary envelope which contains fully the entire CME.  This restricts the volume of space in which the inverse reconstruction can place material, while freely allowing the reconstruction to distribute the material within that envelope in order to best match the data as seen.

We reconstruct the CME as a whole using 3 different constraining envelopes, shown in Figure \ref{FigureFMs}:

a) 'FM series' uses a series of stepwise flux rope forward models manually
fit to time pair

b) 'FM max' uses a single largest extent forward model (taken as the largest of those from
the single steps above),

c) 'Alt FM' refers to an alternative largest extent model fit using a Powell iteration, that yielded a slightly different placement and extent for the constraining element.

The inversion results did not vary greatly when comparing the choice of
forward model constraining envelopes.  Larger envelopes obviously allow
a solution that has a larger volume, but the center of mass positions
for selected regions do not differ, only the distribution of the material about that center of mass.
Each of the three methods produced
a similar center of mass position, and even testing slight variations in
the extent or placement of the constraining envelope does not alter the
positions of the center of mass, despite distributing the solution
into a slightly different volume.  
Our detailed analysis uses the 'FM Max' case with
the largest extent forward model.

Analyzing very
fine structure using this method would be biased in cases where
the structure desired conflicts with the size of the envelope.
For example, investigation of edge densities would show a
dependence on the envelope choice because the envelope can affect the compactness
of the solution.  Since the envelope only constrains the z-axis, we can
separate the CME into horizonal slices without worrying about the choice of
envelope biasing the results.  We split the CME into an 'upper' and 'lower'
segment along $-18\mydeg$ ecliptic line; defining these regions is independent
of constraining envelope as well.

We binned down the $2048\times2048$ COR2 data to $128\times128$ images and created an underlying density cube of size $128^{3}$.  Given the COR2 field of view (FOV) of $30 \Rsun$ at 1AU, the per-pixel FOV $0.234 \Rsun/pixel$ corresponds to a voxel size of $0.234 \Rsun^{3}$.  We limit this analysis to aggregates of voxels rather than individual voxels, since at a separation angle of $44\mydeg$ the problem is still slightly underconstrained.

For the bulk CME position, we achieve with each of the 3 methods a latitude that is independent (within $0.25\mydeg$) of the specific constraining forward model chosen, Table \ref{T-loc}.  The choice of method did result in slightly larger deviations for computed center of mass longitude than for latitude, varying by $1.5\mydeg$ depending on the specific forward model chosen as a constraint.

While the difference images provide the whole observed CME mass, the running difference images highlight the more rapidly changing material, including the CME front.  Comparing both the difference and running difference reconstructions, the center of mass longitudes were comparable, but there was a stronger ($4\mydeg$) difference in the latitude of the CME path between the difference and running difference results.

\begin{table}
\caption{CME location determined from the difference and running difference hybrid reconstructions.  Error brackets for the position show the effect of different assumed forward models.  Latitudes and Longitudes are in the Heliospheric Aries Ecliptic reference frame.  Carrington longitudes are also provided.}
\label{T-loc}
\begin{tabular}{ccccc}
\hline
Location & Source & HAE Lat. & HAE Long.  & Carrington Long.\\
\hline
STEREO B & COR2-A & $0.29\mydeg$ & $76.1\mydeg$ & $310.4\mydeg$ \\
STEREO A & COR2-B & $-0.125\mydeg$ & $120.1\mydeg$ & $354.1\mydeg$ \\
Bulk CME & diff. image & $-18.5\pm0.25\mydeg$ & $11\pm1.5\mydeg$ & $244.5\pm1.5\mydeg$\\
CME front & running diff. & $-14.25\pm0.25\mydeg$ & $11\pm1.5\mydeg$ & $244.5\pm1.5\mydeg$ \\
\hline
\end{tabular}
\end{table}

Analyzing the time series of reconstructions created using a series of expanding self-similar envelopes as constraints yields similar results as using a single maximum extent envelope.  As a preliminary conclusion, the similarity suggests that self-similar expansion is a valid assumption; the constraining volume of the CME increases over time but does not greatly vary in shape, even as the material within that volume shows asymmetries.

\section{Results}
    \label{S-Results}

\subsection{Mass Estimates}
	\label{S-Mass}

The hybrid reconstruction using white light images gives us a measure of the 3D electron density, and to convert to total CME mass we assume the material is fully ionized hydrogen plus 10\% helium.  We reconstruct masses for each voxel and sum voxels to get the total mass for any defined 3D region.  For the total mass, we simply sum all voxels.  Results are averaged across the four time steps (02:22, 03:22, 03:52, 04:22) during which the CME was clearly visible within at least one field of view and no strong data artifacts were present.

We did not enforce mass conservation across each reconstruction time step.  Reconstruction is a statistical minimization, the data has noise, and the constraints are slightly different depending on the fraction of total mass visible in A or B for any given image pair.  Therefore, our result will vary across time steps, and we provide a measure of this in terms of the variability in the mass estimates for the total CME.

We discard the mass estimate from the first time step, 01:52, in our error calculation (though we include it in the plot), as the CME had large portions not visible in either frame and therefore skews the results.
We also neglect in this calculation
the time step at 02:52, where the data are inferior due to a reflection artifact as seen back in Figure \ref{FigureEvo}.

Using the best data, mass was conserved within $\pm7.5\%$ across four reconstructions time frames, and we obtain a best mass estimate by averaging those four values for each section.  The total CME mass is $\sim7\times10^{15}g$, split nearly equally along the streamer line with the mass of the upper region at $3.6\times10^{15}g$ and the mass of the lower region at $3.3\times10^{15}g$, all at $\pm7.5\%$.  The reconstructed CME properties are summarized in Table \ref{T-sum}.

\begin{table}[b]
\caption{Summary properties for the 31 Dec 2007 CME.}
\label{T-sum}
\begin{tabular}{lc}
\hline
Mass estimate (total CME mass) & $7\times10^{15} g$ \\
Upper region mass & $3.6\times10^{15} g$ \\
Lower region mass & $3.3\times10^{15} g$ \\
Avg. HAE (ecliptic) latitude & $-16.4\mydeg\pm2.4\mydeg$ \\
HAE (ecliptic) longitude & $11.0\mydeg\pm1.5\mydeg$ \\
Carrington longitude & $244.5\mydeg\pm1.5\mydeg $ \\
\hline
\end{tabular}
\caption{Summary properties of the CME as a whole and for the selected upper and lower halves.  Values are determined from the data as seen at 02:22UT, 03:22UT, 03:52UT and 04:22UT.  Latitude and longitude are measured in Heliospheric Aries Ecliptic; for the time of this observation the equivalent Carrington longitude for the CME is also provided.}
\end{table}

\subsection{Height-Time and Speed-Time Plots}
	\label{S-HTPlots}

\begin{figure}
\centerline{\includegraphics[width=0.75\textwidth,angle=90,clip=]{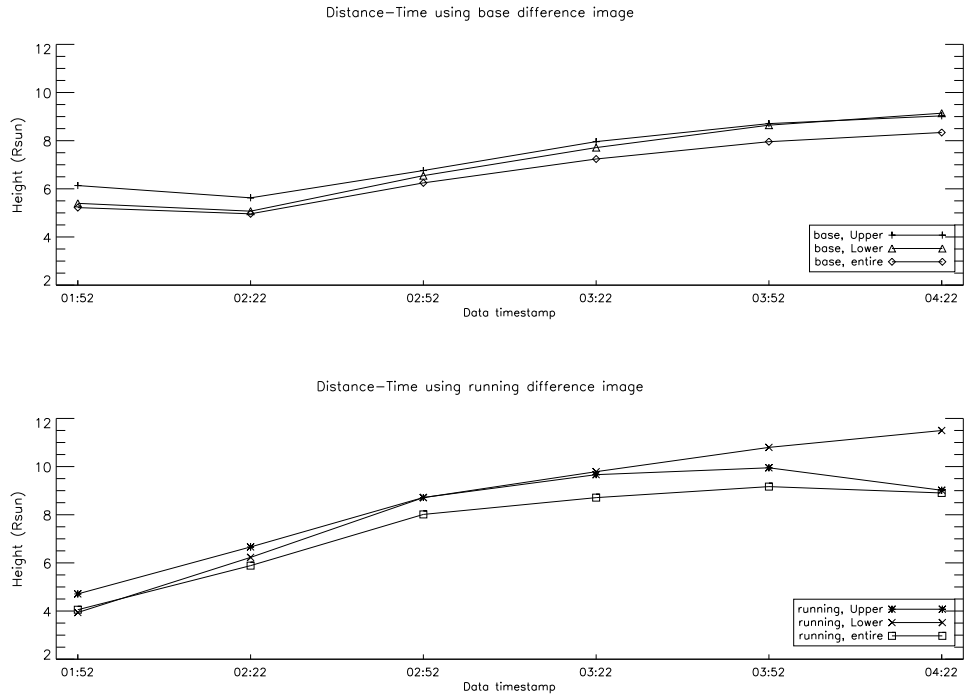}}
\caption{Height-Time plots for the CME for the upper, lower, and entire CME.  The top plot uses base difference images (data frame minus data from 0:22UT), which contain the entire CME.  The bottom plot uses running difference images, which capture just the material that has changed greatly during the 30 minute frame difference.  The running difference contains primarily the rapidly advancing front, while the base difference references the bulk CME mass.  The last point, at 4:22, of both plots is slightly suspect as the upper portion of the CME has begun moving out of the field of view.}
\label{FigureDT}
\end{figure}

The front of the 31 Dec 2007 CME is highly distorted as it travels across the COR2 field of view (2-15 $R_{\odot}$).  Purely geometric projection effects of a flux rope model are not sufficient to reproduce this shape and we contend it is intrinsically distorted or disrupted so as to show the chevron shape of upper lobe, sparse middle, and lower lobe.  In these ranges, the CME is expanding.
Figure \ref{FigureDT} shows the ``true'' Height-Time profile and true space speed for this CME.  The true space speed is taken from the radial distances from the Sun as derived from the 3D reconstruction.
The running difference results, also in Figure \ref{FigureDT}, emphasizes the faster-moving material and therefore weight the front shock more heavily, with a more rapid movement compared to the bulk of the CME.

\begin{figure}
\centerline{\includegraphics[width=0.75\textwidth,angle=90,clip=]{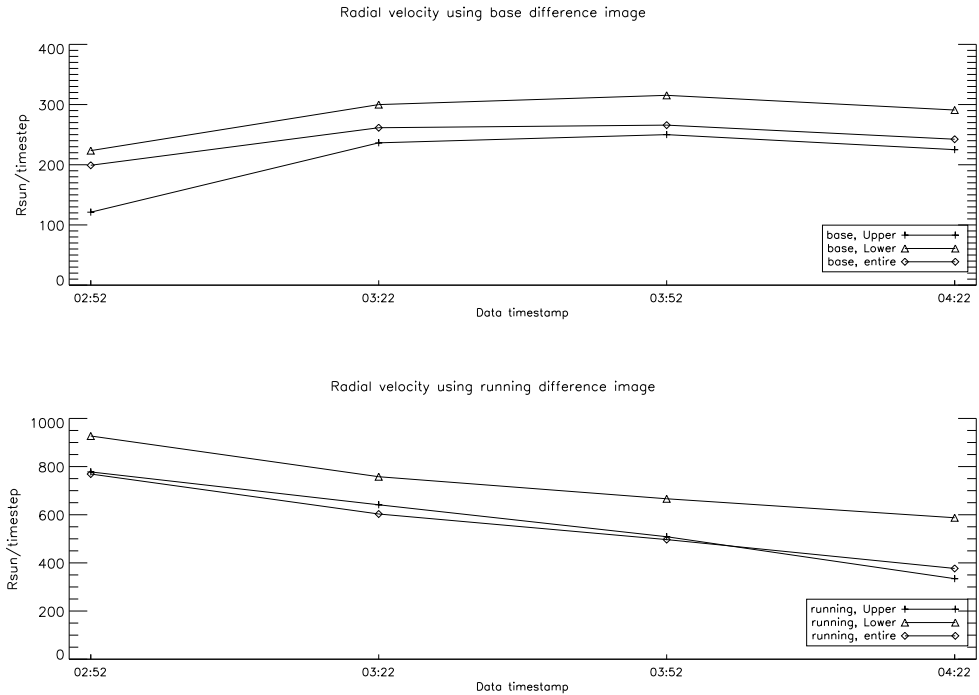}}
\caption{Radial velocities for the CME for the upper, lower, and entire CME.  The top plot uses the base difference images (data frame minus data from 0:22UT), which contain the entire CME.  The bottom plot uses the running difference images, which capture just the material that has changed greatly during the 30 minute frame difference.  The running difference contains primarily the rapidly advancing front, while the base difference references the bulk CME mass.  Note that velocities are vector quantities, so the velocity of the center of mass of the entire object is the 3D vector sum and not simply the numerical average of the upper plus lower velocity.}
\label{FigureKin}
\end{figure}

We plot the radial velocities for the CME, similar to the Height-Time plots, in Figure \ref{FigureKin}.  The bulk CME motion (for the entire CME and for each segment) has an initial acceleration then is relatively flat at approximately $250-300 \kms$.  The more rapidly changing material, as estimated from the running difference images, has an initial CME speed of $750-950 \kms$.  As the CME spreads out, the running difference measure converges to the slower bulk CME speed.  The upper and lower portions of the CME have different velocities and slightly different directions, reinforcing our contention that a single CME speed measure does not suffice in describing the propagation of the CME outwards.

\subsection{Expansion Rate}
	\label{S-Expansion}

The deviation of the base difference and running difference indicates qualitatively the expansion rate of the CME.  However, we note that running difference images include material through the entire CME and therefore provide a slower true space speed than visual tracking of just the shock front.

\begin{figure}
\centerline{\includegraphics[width=0.6\textwidth,angle=90,clip=]{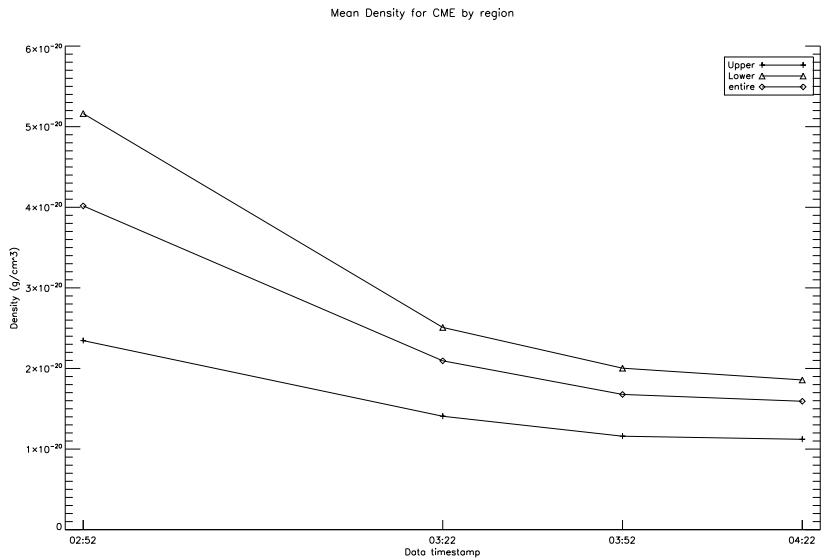}}
\caption{
Mean density evolution over time for the CME in total, as well as for the upper and lower portions individually. Note the reconstruction for time 02:52 is excluded because a data
artifact resulted in a result that did not accurately conserve mass
relative to the other frames.}
\label{FigureDensity}
\end{figure}

To measure the expansion rate of the CME, we extracted the density of occupied voxels (those voxels containing material at or above a 1-sigma standard deviation of the spatial mean).  Within occupied voxels, we calculate a mean density for that segment, plotted in Figure \ref{FigureDensity}.  Although the CME bulk centers of mass are advancing at a similar rate, the upper portion of the CME shows clearly an increasing large expansion, and corresponding lower average density, despite a slightly slower velocity (Figure \ref{FigureKin}).

We suggest that both the upper and lower portions have similar masses and initial motion, and the asymmetries seen are not purely projection effects but reflect the inner heliosphere medium the CME is moving into.  The upper and lower regions are split along the line located along the pre-CME streamer.  The upper portion of the CME is directed along the ecliptic plane and therefore
is the portion of the CME that will first arrive at 1AU.  The lower portion, due to the CME latitude ($-16\mydeg$) and the streamer separation, can be treated as a separate entity possessing the same initial velocity impulse as the upper portion but moving into a different medium and therefore evolving differently.

\section{Conclusion}
    \label{S-Conclusion}

Hybrid reconstruction provides mass, position and density information for a CME viewed from multiple viewpoints.  In contrast to reconstruction methods that apply a monolithic single model or track only a specific feature such as a leading edge, we derive the spatial extent and kinematics of the CME as a whole, and for the separate upper and lower halves, in order to track individual components of a CME.
Additional selection of sub-regions of the CME can be analyzed directly from
the density cubes created by reconstruction. However, sub-regions should be defined based not on arbitrary geometric positions but on whether they are bonafide physical congregations of material.  We restricted ourselves to analyzing the upper and lower lobes as they were the most prominent asymmetrical features.

For the 31 Dec 2007 CME event we studied, the 'chevron' shape of the CME, with the bright observed upper and lower lobes and fainter middle region, is intrinsic to the CME and not just a projection effect.  The evolution of the upper and lower regions are distinct enough that they should be modeled as separate entities. The upper region is initially moving faster but has more deceleration than the lower region. While the estimated masses of the upper and lower regions are very similar (at roughly 50\% the total CME mass), the upper region has a consistently lower reconstructed density and therefore occupies a larger volume of space than the lower region.  The expansion of both regions maintain self-similar shapes but are sufficiently distinct from each other that the rates of their expansion differ.

Some CMEs (e.g. 12-Dec-2008, paper in preparation) have a strongly symmetric shape throughout their propagation, while others such as this 31-Dec-2007 event show asymmetry increasing as they propagate.  The cause of such chevron shapes is not yet understood but such inhomogeneities have been seen since the SOHO era \cite{Wang98}.

Comparison of the true space speed of the two portions during the COR2 observations supports the contention that CME evolution in the inner heliosphere can be expressed as self-similar, shape preserving expansion into an interplanetary medium of differing density \cite{Krall07}.  The upper and lower portions had an identical initial impulse, substantiated by the strongly symmetric shape of the CME at distances closer to the Sun, which develops into a 'chevron' shape as time evolves.  Exact symmetry between the upper and lower regions depart over time as the top is rapidly becoming more rarefied as the CME evolves.  Such asymmetry should be considered when determining the true space speeds and evolution of the CME at increasing distances from the Sun.



\begin{acks}
The authors thank the STEREO SECCHI consortia for supplying their data. 
Portion of this work were funded by the SECCHI mission.
The SECCHI data used here were produced by an international consortium
of the Naval Research Laboratory (USA), Lockheed Martin Solar and
Astrophysics Lab (USA), NASA Goddard Space Flight Center (USA),
Rutherford Appleton Laboratory (UK), University of Birmingham (UK),
Max-Planck-Institut for Solar System Research (Germany), Centre
Spatiale de Lie`ge (Belgium), Institut dâOptique TheÂ´orique et
AppliqueeÂ´ (France), and Institut dâAstrophysique Spatiale (France).W

This research was performed while A. Antunes held a National Research Council Research Associateship Award at Naval Research Laboratory.
\end{acks}



\end{article} 
\end{document}